\documentclass[apj,iop]{emulateapj}
\usepackage{apjfonts}
\usepackage{graphicx}
\usepackage{amsmath,amsthm,amsfonts,amssymb,extarrows}
\usepackage[breaklinks,colorlinks,urlcolor=blue,citecolor=blue,linkcolor=blue]{hyperref}
\usepackage{natbib}
\shorttitle{F\lowercase{e}~\textsc{xvii} $n=$3$\rightarrow$2 transitions}
\shortauthors{Shah et al.}

\begin{document}

\title{{Revisiting the F\lowercase{e}~\textsc{xvii} line emission problem: \\Laboratory measurements of the 3\textit{\lowercase{s}} -- 2\textit{\lowercase{p}} and 3\textit{\lowercase{d}} -- 2\textit{\lowercase{p}} line-formation channels}}

\author{Chintan Shah\altaffilmark{1, $\ast$}, Jos\'e R. Crespo L\'opez-Urrutia\altaffilmark{1}, Ming Feng Gu\altaffilmark{2}, Thomas Pfeifer\altaffilmark{1}, Jos\'e Marques\altaffilmark{3, 4}, Filipe Grilo\altaffilmark{4}, \\Jos\'e Paulo Santos\altaffilmark{4}, and Pedro Amaro\altaffilmark{4, $\star$} }

\affil{$^1$Max-Planck-Institut f\"{u}r Kernphysik, D-69117 Heidelberg, Germany}

\affil{$^2$Space Science Laboratory, University of California, Berkeley, CA 94720, USA}

\affil{$^3$University of Lisboa, Faculty of Sciences, BioISI - Biosystems \& Integrative Sciences Institute, Lisboa, Portugal}

\affil{$^4$Laborat\'{o}rio de Instrumenta\c{c}\~{a}o, Engenharia Biom\'{e}dica e F\'{\i}sica da Radia\c{c}\~{a}o (LIBPhys-UNL), \\Departamento de F\'{\i}sica, Faculdade de Ci\^{e}ncias e Tecnologia, FCT, Universidade Nova de Lisboa, 2829-516 Caparica, Portugal}

\altaffiltext{$\ast$}{\href{mailto:chintan@mpi-hd.mpg.de}{chintan@mpi-hd.mpg.de}}
\altaffiltext{$\star$}{\href{mailto:pdamaro@fct.unl.pt}{pdamaro@fct.unl.pt}}

\begin{abstract}

We determined relative X-ray photon emission cross sections in Fe~\textsc{xvii} ions that were mono-energetically excited in an electron beam ion trap. 
Line formation for the $3s$ ($3s-2p$) and $3d$ ($3d-2p$) transitions of interest proceeds through dielectronic recombination (DR), direct electron-impact excitation (DE), resonant excitation (RE), and radiative cascades. 
By reducing the electron-energy spread to a sixth of that of previous works and increasing counting statistics by three orders of magnitude, we account for hitherto unresolved contributions from DR and the little-studied RE process to the $3d$ transitions, and also for cascade population of the $3s$ line manifold through forbidden states. 
We found good agreement with state-of-the-art many-body perturbation theory (MBPT) and distorted-wave (DW) method for the $3s$ transition, while in the $3d$ transitions known discrepancies were confirmed.  
Our results show that DW calculations overestimate the $3d$ line emission due to DE by $\sim$20\%. 
Inclusion of electron-electron correlation effects through the MBPT method in the DE cross section calculations reduces this disagreement by $\sim$11\%.
The remaining $\sim$9\% in $3d$ and $\sim$11\% in $3s/3d$ discrepancies are consistent with those found in previous laboratory measurements, solar, and astrophysical observations.
Meanwhile, spectral models of opacity, temperature, and turbulence velocity should be adjusted to these experimental cross sections to optimize the accuracy of plasma diagnostics based on these bright soft X-ray lines of Fe~\textsc{xvii}.

\end{abstract}


\maketitle

%
%-----------------------------------------------------------------------------
%
\section{Introduction}
%
%-----------------------------------------------------------------------------
%
%
{X-rays from astrophysical hot plasmas} at a few MK are recorded by grating spectrometers onboard the~\textit{Chandra} and~\textit{XMM-Newton} X-ray observatories.  They are dominated by the $L$-shell $3d-2p$ and $3s-2p$ transitions in the 15--17~\AA~range from Fe~\textsc{xvii} (Ne-like ions)~\citep{pfk2003,chd2000} that are used for electron temperature, density ~\citep{mrd2001, behar2001, xpb2002, beiersdorfer2018}, velocity turbulence, and X-ray opacity diagnostics~\citep{brickhouse2005, wzc2009, sfs2011, pzw2012, kem2014}.  
Decay from the states $[2p_{1/2}^5 3d_{3/2}]_{J=1}$, $[2p_{3/2}^5 3d_{5/2}]_{J=1}$, and $[2p_{3/2}^5 3d_{3/2}]_{J=1}$ to the $[2p^6]_{J=0}$ ground state produces the $3d-2p$ transitions called $3C$, $3D$, and $3E$, respectively. 
The $3s-2p$ decays known as $3F$, $3G$, and $M2$ lines proceed from $[2p_{1/2}^5 3s_{1/2}]_{J=1}$, $[2p_{3/2}^5 3s_{1/2}]_{J=1}$, and $[2p_{1/2}^5 3s_{1/2}]_{J=2}$, also to $[2p^6]_{J=0}$.

\begin{figure*}[t]
	\centering
	\includegraphics[width=\textwidth]{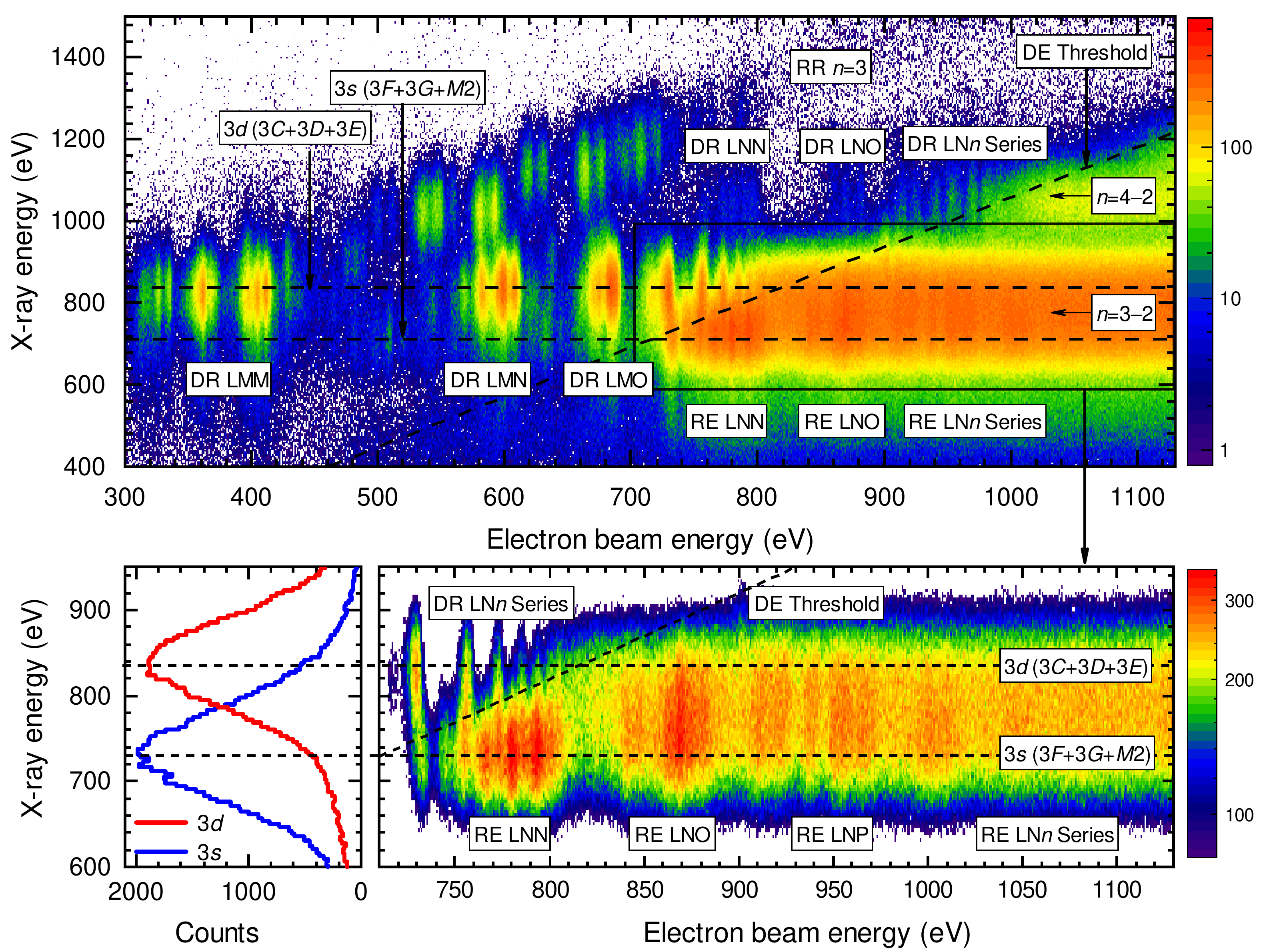}
	\caption{\label{fig:spectro} Measured X-ray flux in dependence of photon (3.7 eV/bin) and electron energy (0.9 eV/bin). Bright DR and RE resonances are labeled by their respective dielectronic capture channel and can contribute as unresolved satellites to the $3d$ $(3C+3D+3E)$ and $3s$ $(3F+3G+M2)$ transition groups marked by horizontal dashed lines. The bottom-right plot highlights distinct RE features on the top of the DE band and the bottom-left plot shows a clear separation of $3s$ and $3d$ line manifolds in the regions where they do not overlap.}
\end{figure*}

Since the optically thick line $3C$ and the intercombination line $3D$ both have low contributions from cascades~\citep{mrd2001}, their intensity ratio mainly depends on direct electron-impact excitation (DE) and dielectronic recombination (DR) processes.
The diagnostic utility of this ratio is strongly hampered by discrepancies between observations and various predictions of their oscillator and collision strengths, which can lead to overestimating opacity effects~\citep{par1973, bbl1998, lkt2000, bbb2002, bbg2004, bbc2006, che2007, chen2008, lib2010}. 
Although laboratory measurements shown to be in agreement with each other~\citep{bbc2006,glt2011}, they disagree with state-of-the-art predictions of the $3C/3D$ collision-strength ratio (see~\citet{brown2012}). 
So far, only one theoretical work~\citep{chen2008mnras} predicted a low $3C/3D$ ratio in agreement with experimental data~\citep{bbc2006,glt2011}. However, a latter calculation performed by the same author~\citep{chen2011} produced a high $3C/3D$ ratio, in consensus with the most-advanced theoretical predictions~\citep{safronova2001,gu2009,sbr2012}. 
Furthermore, a comprehensive comparison of the $3C/3D$ ratio for the Ne-like isoelectronic sequence from Ar~\textsc{ix} to Kr~\textsc{xxvii} exhibited discrepancies between measurements and predictions~\citep{slt2015}. Particularly, theories deviate by 10--35\% from experiments in the case of Fe~\textsc{xvii}. 
A novel X-ray laser spectroscopy measurements at the LCLS free-electron laser facility measured $3C/3D$ oscillator-strength ratio, which is also proportional to the collision-strength ratio under Bethe approximation, also produced a low $3C/3D$ ratio in agreement with previous measurements and observations (see~\citet{sbr2012} and its supplementary material). 
Nonlinear effects in X-ray lasers could reduce the $3C/3D$ ratio~\citep{ock2014, lbl2015}. While a recent semi-empirical calculation has predicted a value close to that experimental result~\citep{mba2017}, this theoretical approach was quickly disputed~\citep{wpe2017}.
In short, in spite of forty years of efforts, the $3C/3D$ intensity-ratio discrepancy remains essentially unsolved to this date.

Alternative approaches use the forbidden $M2$ line or the $3G+M2$ line blend for opacity and turbulence-velocity diagnostics since the $M2$ line is optically thinner than the $3D$ line~\citep{pzw2012,wzc2009}.
However, astrophysical observations of the $(3G+M2)/3C$ ratio and $3s/3d$ or $(3F+3G+M2)/(3C+3D+3E)$ ratio also depart from theories and spectral models~\citep{pzw2012}, 
and laboratory ratios are consistently larger than the calculated ones~\citep{bbb2002,bbg2004}. 
This could be explained if $M2$ is mainly fed by complex cascades following DE, with contributions from resonant excitation (RE)~\citep{beiersdorfer1990,doron2002,fgu2003,tsuda2017} process. Another argument points to the same cause than the $3C/3D$ discrepancy.  

Understanding whether $3C/3D$ and $3s/3d$ discrepancies are caused by DR, RE, or cascade contributions to $3s$ and $3d$ line manifolds is highly critical given wide-reaching diagnostic applications of these lines. 
Most previous laboratory works mainly focused on DE cross section measurements at discrete electron beam energies.  
However, comprehensive laboratory validations of the DR, DE, and RE contributions to these transitions are essential to construct reliable spectral models, and urgently needed in view of the upcoming high-resolution space missions {\it XRISM Resolve}~\citep{xrism2018}, {\it Arcus}~\citep{arcus}, and {\it Athena}~\citep{barret2016}.

Here we perform laboratory measurements of differential cross sections in the 0.3--1.1~keV energy range of the Fe~\textsc{xvii} $3s$ $(3F+3G+M2)$ and $3d$ $(3C+3D+3E)$ emissions that are driven by DR, DE, and RE processes. 
Using an electron beam ion trap (EBIT), we produced an ion population mainly consisting of Fe~\textsc{xvii} ions and scanned the electron beam energy with an energy spread of only 5~eV full-width-at-half-maximum (FWHM). This, together with three orders of magnitude higher counts compared to previous experiments, allow us clearly to distinguished narrow RE and DR features from DE ones as a function of electron beam energy. 
Furthermore, we perform calculations of $3s$ and $3d$ line emission cross sections using both state-of-the-art distorted-wave (DW) and many-body perturbation theory (MBPT) methods. 
We found overall agreement of the experimental $3s$ cross sections with both DW and MBPT predictions, thus benchmarking the DE cross sections with respective cascades and the strong RE features. 
On the other hand, we observe discrepancies between both theories and the experimental $3d$ cross section, consistent with previous laboratory measurements, solar, and astrophysical observations of $(3G+M2)/3C$ ratio and $3s/3d$ or $(3F+3G+M2)/(3C+3D+3E)$ ratios. 
Our results indicate that a fundamental discrepancy in $3d$ formation likely causes the reported inconsistencies of models with laboratory and astrophysical observations.

\begin{figure*}[t]
	\centering
	\includegraphics[width=\textwidth]{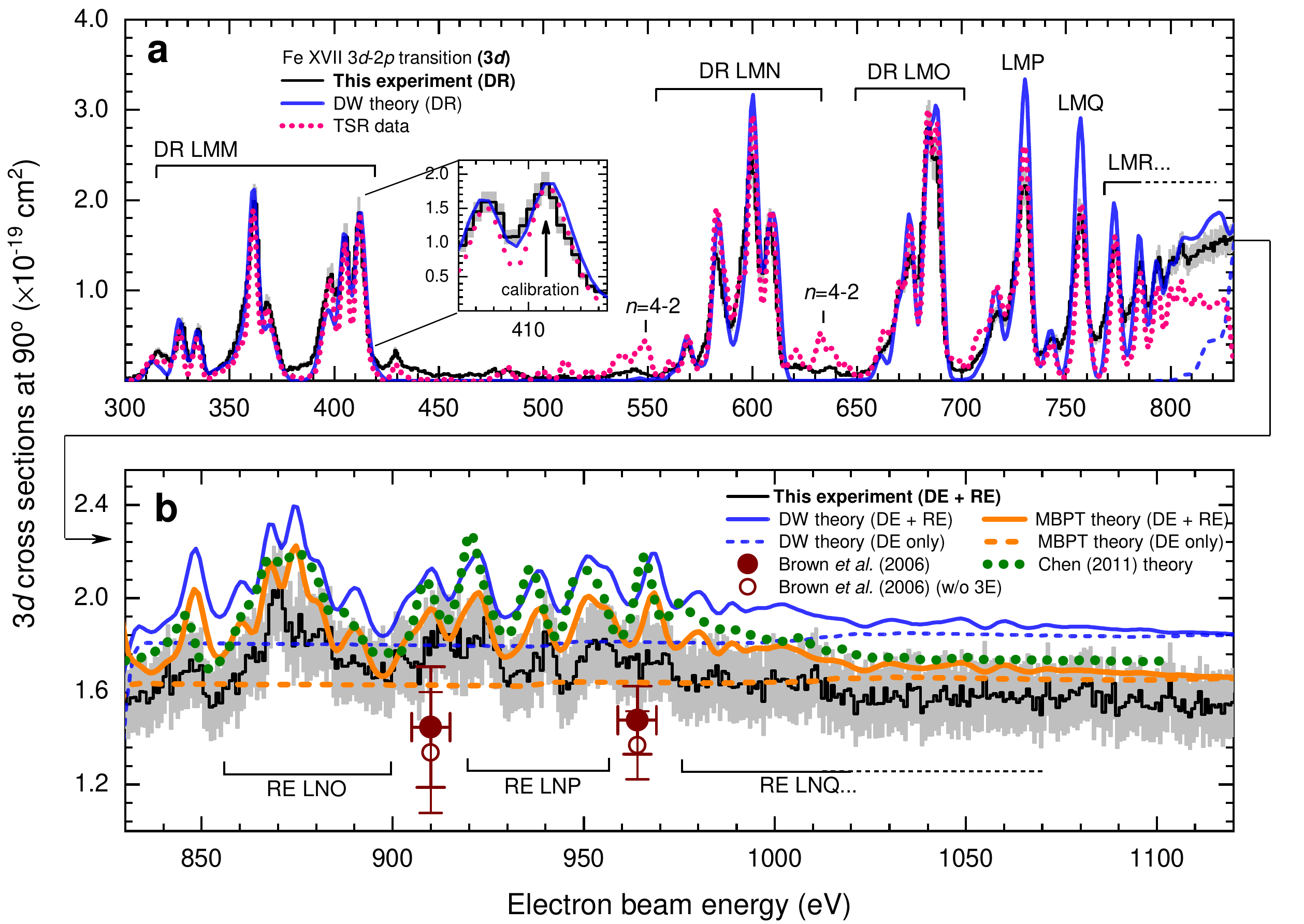}	
	\caption{
		\label{fig:3d} Black curves: Experimental cross sections observed at 90$^\circ$ vs.~electron beam energy for the $3d$ $(3C+3D+3E)$ manifold. Grey band: Total ($1\sigma$ statistical + systematics) uncertainty. 
		(a) Measured DR cross sections below the DE threshold. Dotted magenta curve: \textit{Independently} calibrated high-resolution results from the Test Storage Ring (TSR) in Heidelberg~\citep{schmidt2009} that were folded with a 5\,eV Gaussian for comparison. Blue solid curve: our FAC-DW predictions. Inset: DR resonance used for normalization of the 2D map to our FAC results (Fig.~\ref{fig:spectro}).  
		(b) DE cross sections including above-threshold RE peaks. FAC-DW predictions: Blue solid (DE + RE) and dashed (DE only) curves. FAC-MBPT predictions: Orange solid (DE + RE) and dashed (DE only) curves. Green dotted curve: Breit-Pauli $R$-matrix predictions of~\citet{chen2011}. Solid circles: measured LLNL-EBIT cross sections~\citep{bbc2006}. For comparison with~\citet{bbc2006}~and~\citet{chen2011} data, $3E$ cross sections are added from our FAC DW theory. Hollow circles:~\citet{bbc2006} data without addition of the 3E line ($\sim$7\%). The data behind the figure are available in the machine-readable table.
	} 
\end{figure*}

%-----------------------------------------------------------------------------
%
\section{Experimental technique}
%
%-----------------------------------------------------------------------------
%
We used FLASH-EBIT~\citep{epp2007,epp2010} at the Max-Planck-Institut f\"ur Kernphysik in Heidelberg to produce Fe~\textsc{xvii} ions from a molecular beam of iron pentacarbonyl. The monoenergetic electron beam induces a negative space-charge potential that radially traps the ions; in the axial direction, a set of cylindrical drift tubes electrostatically confines them. 
In the trap, the electrons have a well defined kinetic energy due to the acceleration potentials corrected by the space-charge contributions of the electron beam and trapped ions. 
Collisions between beam electrons and trapped ions efficiently drive ionization, excitation, and recombination processes. 
The generated X-rays are registered at 90$^\circ$ to the beam axis with a silicon-drift detector (SDD) with a photon-energy resolution of $\approx$120\,eV FWHM at 6\,keV, which separate transitions from the $3s$ and $3d$ manifolds, see the bottom-left inset of Fig.~\ref{fig:spectro}.

To maximize Fe~\textsc{xvii} purity, a saw-tooth scan consists of a breeding time of 0.5~s at 1.15~keV beam energy (below the Fe~\textsc{xvii} ionization threshold at 1.26~keV), followed by a 40\,ms-long energy scan from 0.3~keV to 1.1~keV during which only a small fraction of the ions recombine. This method efficiently suppressed lower charge states, as only very faint $LMM$ DR resonances from Fe~\textsc{xv-xvi} are obtained. Measurements at breeding energy of 0.5~keV, just above the Fe~\textsc{xvi} ionization threshold allowed us to separately identify those weak contributions. 
The DR paths are named in analogy to Auger nomenclature, e.~g.~, $LMM$ implies the $L\rightarrow M$ resonant excitation due to a free electron recombining into the $M$-shell.
Moreover, the spectra taken on the upward and downward energy scans show differences of less than 2\% and confirm a nearly constant population of Fe~\textsc{xvii} ions. %, thus facilitating the determination of energy-dependent cross sections. 

During the fast energy scans, the beam current $I_e$ was also adjusted synchronously with $E$ to keep the electron density $n_e\propto I_e/\sqrt{E} $ constant~\citep{sbk2000}, minimizing a modulation of the space-charge potential, and concomitant heating that causes ion losses. 
The trap was emptied every 5~s to prevent accumulation of unwanted (Ba, W) ions emanating from the electron-gun cathode.

\begin{figure*}[t]
	\centering
	\includegraphics[width=\textwidth]{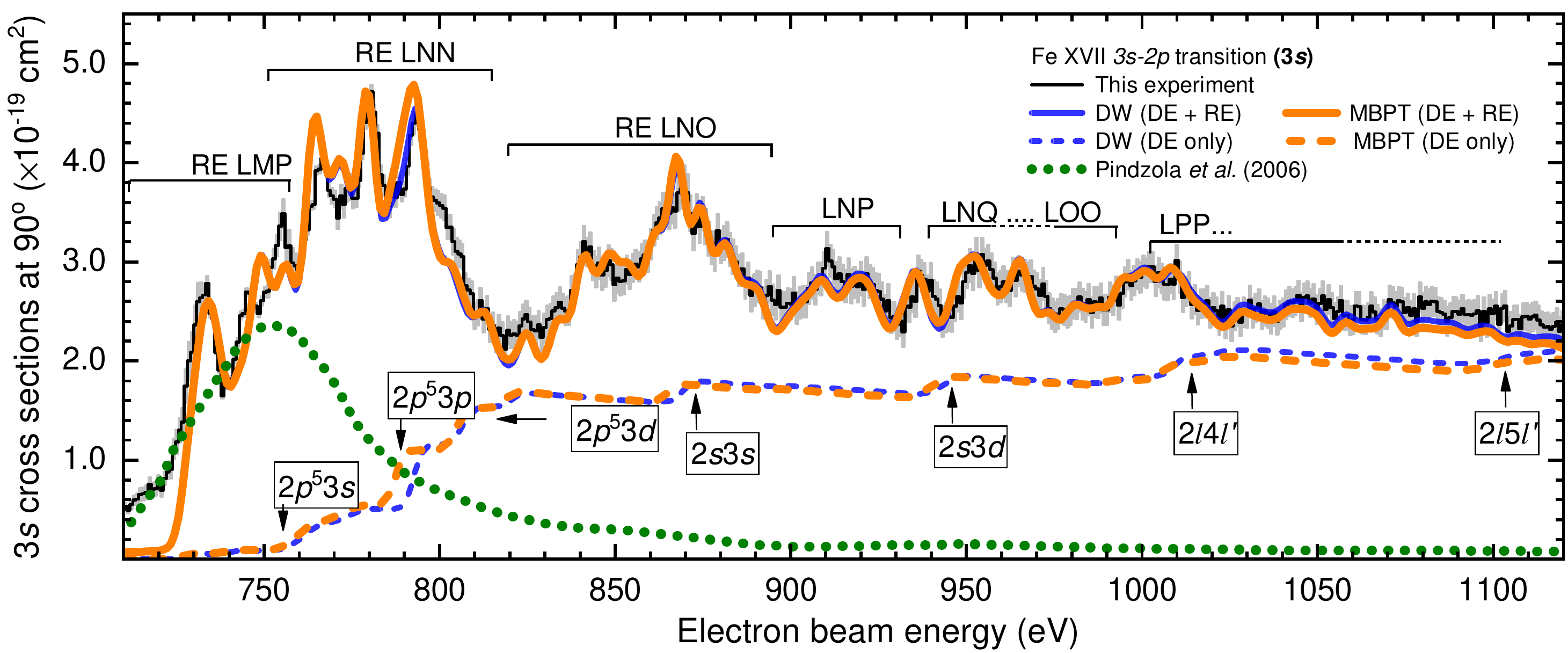}	
	\caption{
		\label{fig:3s} Same as in Fig.~\ref{fig:3d}b, except for the $3s$ $(3F+3G+M2)$ manifold. Green dotted line: 30-eV brodened $2p^53s$ cross sections presented by~\cite{pindzola2006}.
	}
\end{figure*}

In this work, we reached an electron-beam energy spread of $\sim$5 eV FWHM at 800~eV, a six-to-ten-fold improvement over previous works~\citep{bbc2006,glt2011,beiersdorfer2017} in this energy range by an application of forced evaporative cooling technique~\citep{penetrante1991}. 
Here, the cooling of ions optimized by lowering as much as possible the axial trapping potential. 
Within the radial trapping potential generated by the space charge potential of the electron beam, the cooled ions are more concentrated at the bottom, thus reducing the energy dispersion due to the space charge.  
The details of this technique are discussed in~\citet{beilmann2009,beilmann2010,beilmann2011,sas2016,sas2018,micke2018}. 
Since the ion trapping parameters were kept constant during the electron beam energy scan, this technique does not introduce any systematic effects on the present measurements.

%
%-----------------------------------------------------------------------------
%
\section{Theory}\label{sec:th}
%
%-----------------------------------------------------------------------------
%

Within an independent resonance approximation model, the DR and RE processes can be described as a two-step process.
First, a doubly excited state is formed by dielectronic capture (DC), i.e., a capture of a free electron by an ion with excitation of a bound electron. Here, we focus on the $L$-shell. 
While in DR this excited state decays radiatively, RE includes an autoionization process which leaves a $L$-shell hole that relaxes by photon emission. 

We used the parallel version of Flexible Atomic Code\footnote{\href{https://github.com/flexible-atomic-code/fac}{\texttt{https://github.com/flexible-atomic-code/fac}}} (FAC v1.1.5)~\citep{gu2008} to compute the electronic structures of Fe~\textsc{xvii-xvi} ions. We evaluated DR, DE, and RE with extended sets of configurations, full-order configuration mixing, and Breit interaction~\citep{ass2017, sas2018}. 
For DC channels of both DR and RE, we included $2s^22p^5nln'l'$ and $2s 2p^6 nln'l'$ configurations.
All the DR radiative paths of these states were included ($2s^22p^6n’l'$), which corresponds to the main radiative recombination (RR) paths. 
Additionally, all Auger paths addressing RE are taken into account ($2s^22p^6n'l'$ and $2s 2p^6 n'l'$). They are also the main paths of DE. 
In all sets of configurations, we included principal quantum numbers and orbital angular momentum quantum numbers of $n\le7$, $n'\le60$ and $l,l'<8$ that resulted into roughly half-a-million eigenstates in our calculations. 
The output was fed into the collisional-radiative model of FAC~\citep{gu2008} for obtaining steady-state populations of Fe~\textsc{xvii-xvi} states connected by the aforementioned processes and radiative cascades. 
It solves a system of coupled rate equations on a fine grid of electron beam energies and electron density matching the experimental conditions. 
The resulting level populations are used to calculate the line emission for comparison with the measurement.

For the DE cross sections, we used the MBPT implementation of FAC. 
The treatment of DE with the MBPT method is described by~\citet{gu2009}. 
In essence, a combined configuration interaction and second-order MBPT method~\citep{gu2006} is used to refine the energy levels and multipole transition matrix elements. 
Collision strengths of DE under the DW approximation can be split into direct and exchange interaction contributions. 
Allowed transitions such as $3C$ and $3D$ lines of Ne-like ions have a dominant contribution of the direct interaction that is proportional to the corresponding matrix elements. 
We, therefore, use the MBPT-corrected matrix elements to scale the direct contribution to the DW collision strength and obtain correlation corrections to the DE cross sections.

Due to the unidirectional electron beam and the side-on X-ray observation, both polarization and emission anisotropy must be accounted for~\citep{beiersdorfer1996,shah2015}. 
We used also FAC to calculate the X-ray polarization following DR, DE, and RE. 
Depolarization due to radiative cascades and cyclotron motion of the electrons was taken into account~\citep{beiersdorfer1996,gu1999}.
The transversal electron energy component due to the cyclotron motion of electrons inside the electron beam was estimated in our previous work~\citep{sas2018}. For the present experimental conditions, the depolarization due to this essentially resulted in no change in the total X-ray polarization. 
	
All theoretical data on cross sections and polarization presented in Figs.~\ref{fig:3d},~\ref{fig:3s},~and~\ref{fig:ratio} are listed in the accompanying machine-readable table.

%-----------------------------------------------------------------------------
%
\section{Data Analysis}
%
%-----------------------------------------------------------------------------
%

Figure~\ref{fig:spectro} shows the X-ray intensity as a function of the electron beam (abscissa) and the X-ray energies (ordinate). 
Features due to electron recombination can be recognized above and below the DE threshold. 
Above it, we resolve DR channels populating the $3d$ manifold from $LMM$ to $LMT$, while unresolved $LMn$~DR channels up to~$n>60$ pile up near the $3d$ threshold. 
Below the threshold, photon emissions produced after DE of $L$-shell electrons appear as horizontal bright band comprising both the $3d$ and $3s$ manifolds. {Bright spots on top of the continuous DE emission band are RE features, see the bottom-right inset of Fig.~\ref{fig:spectro}.}

Two horizontal regions of interest (ROI) with an X-ray energy width of $\pm$30~eV and centered on the respective line centroids were carefully selected on the 2D map for distinguishing the contributions of the $3s$ and $3d$ manifolds. 
X-ray photon counts within these ROIs were then projected onto the electron-energy axis, see Figs.~\ref{fig:3d}~and~\ref{fig:3s}. 
In front of the windowless SDD detector, a 1$\mu$m carbon foil blocks UV light from the trapped ions. 
At the X-ray energies of interest varying from 650 to 900~eV, the foil has a transmission of 26--52\%. We normalized counts with known transmission coefficients~\citep{hanke1993}. To verify them, we carried out measurements of Ly$\alpha$ and radiative recombination (RR) emissions from O~\textsc{viii} and Ne~\textsc{x}, and found them in agreement within ~3\%.   
Furthermore, X-ray yields were also normalized to the cyclically time-varying electron beam current density.  
After that, we determine the relative differential cross sections. 

Previously,~\citet{bbc2006} determined the $3C$ and $3D$ cross sections using an X-ray microcalorimeter by normalizing their experiment with theoretical RR cross sections. %, and determined the $3C$ and $3D$ cross sections.
This was not possible in the present experiment due to pile-up and contamination of the RR band, since the detector does not resolve RR transitions into the 3$s$, 3$p$, and 3$d$ sub-shells. 
We thus selected a single DR resonance, $[1s^2 2s^2 2p_{1/2} 2p_{3/2}^4 3d_{3/2} 3d_{5/2}]_{J=7/2}$ at $\sim$412 eV electron beam energy for normalizing the $3s$ and $3d$ projections, see the inset of the Fig.~\ref{fig:3d}. 
An~\textit{ab initio}~calculation for this resonance using FAC under DW approximation was employed. 
Its strength can be traced to a single and strong resonance in the $LMM$ channel. 
This channel has the simplest structure of all DR channels and this presumably makes its prediction theoretically more reliable. 
Moreover, theoretical treatments of DR and RE are more reliable than the DE due to the involvement of only one continuum state in DR and RE, in contrast to two continuum states in the DE treatment.  
By fitting the experimental and the theoretical projections (convoluted with a Gaussian function), a normalization factor of $1.42\times10^{21}$~counts per cm$^2$ with a 2\% fitting error was found. 
The complete projection was accordingly rescaled, yielding the differential cross sections for both the $3d$ and $3s$ manifolds at 90 degrees, see Fig.~\ref{fig:3d}.

To verify our normalization factor, we also used a theoretical cross-section value of single RE-$LMP$ resonance at $\sim$734 eV electron beam energy and found a normalization factor of $1.46\times10^{21}$~counts per cm$^2$ with a 3\% fitting error. We select this RE resonance because it is relatively free from the DE background. 
Moreover, we further normalized our experiment with measured cross sections of $3C$ and $3D$ by~\citet{bbc2006}, which are normalized to RR cross sections, at $\sim$964 eV electron beam energy. 
All three normalization procedures show consistency in inferred cross sections within their uncertainties.

We also utilized the DR rate coefficients {\textit{independently}} measured at the Test Storage Ring (TSR) at the Max-Planck-Institut f\"ur Kernphysik in Heidelberg by the merged-beam method. 
There, a cooled Fe~\textsc{xvii} ion beam is brought to overlap with an electron beam as the electron target~\citep{schmidt2009}.
The yield of Fe~\textsc{xvi} ions, not X-ray photons, is detected, resulting in a high-resolution absolute total recombination-rate coefficients with an estimated total uncertainty of 20\%, which we corrected for observations at 90$^\circ$ and convoluted to a 5~eV Gaussian for the comparison.   
In the Fig.~\ref{fig:3d}, the \textit{independently} calibrated TSR data (without any normalization) are shown for the comparison along with the present EBIT data. 
We not only found a very good agreement for the DR resonance we used for the normalization, but also an overall consensus between TSR and EBIT data for all measured DR resonances. 
Moreover, we also compared our inferred cross sections with those measured at the LLNL-EBIT by~\citet{bbc2006} at 910 eV and 964 eV energies. Both EBIT data are in agreement within their reported uncertainties and consistent with or without addition of 3E cross sections. 
This overall agreement of three independent experiments for the $3d$ line manifold is shown in Fig.~\ref{fig:3d}a and~\ref{fig:3d}b. 

\begin{figure*}[t]
	\centering
	\includegraphics[width=\textwidth]{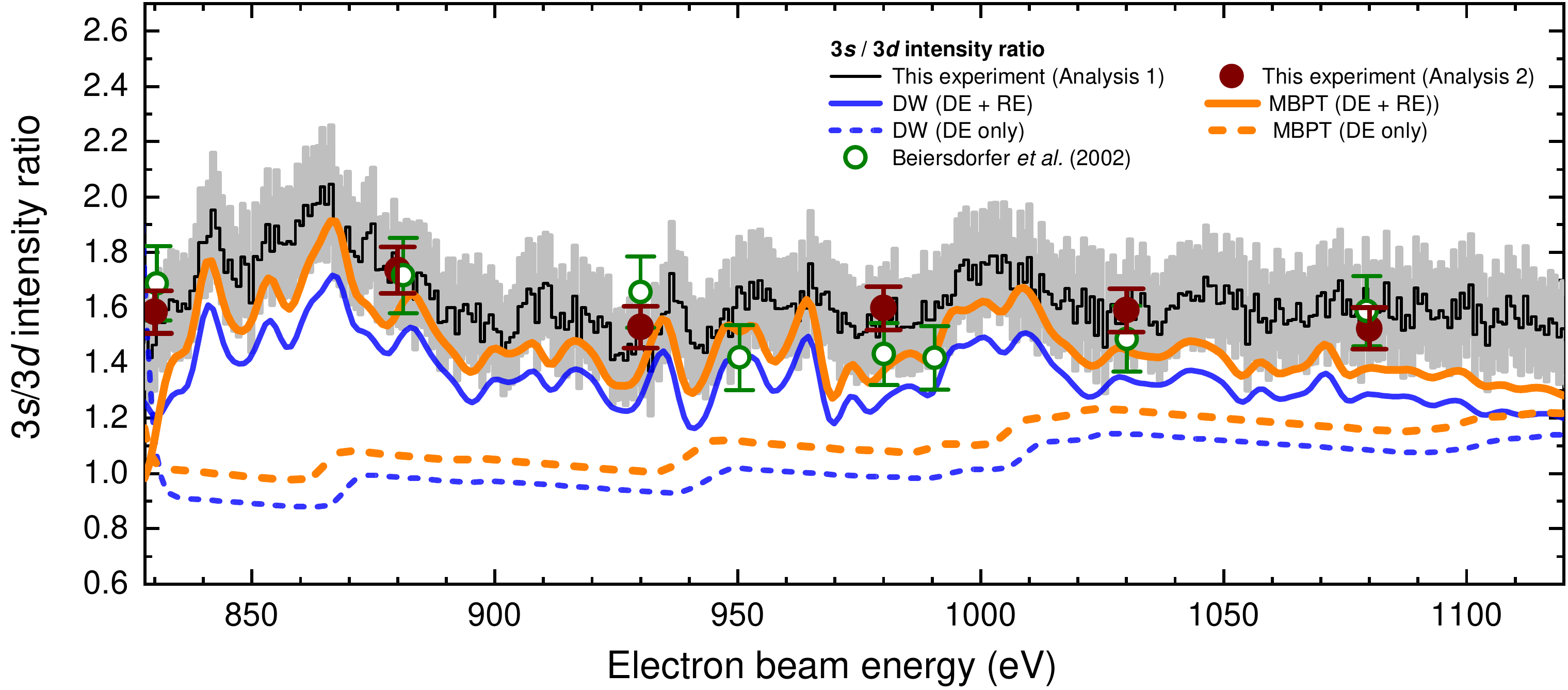}	
	\caption{\label{fig:ratio} Measured $3s$ $(3F+3G+M2)$ to $3d$ $(3C+3D+3E)$ line intensity ratios as a function of the electron beam energy. Black curve: present experiment (based on DR-normalization procedure--Analysis 1). Grey shaded area: $1\sigma$ statistical + systematic error. Solid circles: present experiment (independent of normalization--Analysis 2) Open circles: LLNL-EBIT microcalorimeter~\citep{bbb2002}. Blue solid and dashed curves show distorted-wave (DW) predictions that include (DE + RE) and exclude RE (DE only), respectively. Similarly, orange solid and dashed curves show the many-body perturbation theory (MBPT) predictions.}
\end{figure*}

We further check the effect of the $3s$-$3d$ ROI selection, since even though our detector resolves them, the wings of both transitions partially overlap. 
This is evident in Fig.~\ref{fig:spectro}: the low energy tail of $3d$ peak contaminates the $3s$ manifold.
To check those blends, we shifted the $3s$ and $3d$ ROIs up to $\pm$15 eV from their respective centroids, leading to effects on the normalization factor of 9\%. 
We conservatively add this $\sim$9\% uncertainty to our normalization factor, in addition to the $\sim$2\% uncertainty from the $LMM$ fits. 
Total uncertainty sources are thus: $\sim$2\% from 
statistics, $\sim$3\% from the carbon-foil transmission, and $\sim$9\% from the normalization factor. 
The resulting cross sections with their overall uncertainties of DR, RE, and DE driven $3d$ and $3s$ line emissions are shown in Figs.~\ref{fig:3d}~and~\ref{fig:3s}, respectively, and compared with FAC-MBPT and FAC-DW calculations performed in this work as well as previous works of~\citet{chen2011,pindzola2006}.

The ratios of normalized cross sections of $3s$ and $3d$ line manifold are also shown in Fig.~\ref{fig:ratio} (dubbed as Analysis 1) as a function of electron beam energies. 
These $3s/3d$ ratios are then compared with LLNL-EBIT ratios measured with three different high-resolution X-ray spectrometers~\citep{bbb2002}. 
We found a good agreement with LLNL-EBIT data, however disagreement with a microcalorimeter measurement of~\citet{lkt2000}. 

In the second analysis, we did not rely on our normalization procedure in order to obtain $3s/3d$ line ratios. 
As can be seen in the bottom-left inset of Fig.~\ref{fig:spectro}, we can clearly separate X-ray peaks of $3s$ and $3d$ by selecting a particular ROI (vertical slices) along the electron beam energy axis where both do not overlap, e.g. at 690 eV and at 745 eV for $3d$ and $3s$, respectively.  
By fitting a Gaussian with a low-energy Compton wing, we extracted the line centroids and widths of $3s$ and $3d$ peaks. 
We then selected several other ROIs along the electron beam energy axis with $\pm25$ eV widths, and projected X-ray counts within these ROIs onto X-ray energy axis. 
Here $3s$ and $3d$ peaks are not separated, however, line intensities of both $3s$ and $3d$ manifolds can still be extracted by fitting Gaussians while fixing line centroids and widths to previously-obtained values in the fitting procedure. 
The inferred $3s/3d$ ratios by this method (dubbed as Analysis 2) are shown in Fig.~\ref{fig:ratio}. 
Both analysis methods are in very good agreement with each other as well as with the LLNL-EBIT data of~\citet{bbb2002}. 
This, in turn, reassures our normalization procedure to obtain the $3s$ and $3d$ cross sections. 
We again emphasize that our normalization is based on a single theoretical value of the DR resonance at $\sim$412 eV electron beam energy.

\begin{table*}
	\centering
	\caption{Experimental and theoretical values of the integral cross sections (10$^{-18}$~cm$^2$eV) for the $3s$ and $3d$ manifolds and their dimensionless ratio. The predicted contributions of various processes from DW and MBPT calculations are compared with experimental values.}
	\begin{tabular}{ccccccccccc}
		\hline\hline
		Line  & $E_e$ range & \multicolumn{4}{c}{Integral cross sections (in 10$^{-18}$ cm$^2$eV)} & \multicolumn{2}{c}{\% Contribution of processes} & \% MBPT correction & \multicolumn{2}{c}{\% Required correction} \\
		& (eV)  & This exp. & \multicolumn{2}{c}{DW Theory} & MBPT Theory & \multicolumn{2}{c}{DW theory} & to DW theory & \multicolumn{2}{c}{to theory} \\
		
		&       &       & DE + RE & DE only & DE only & \% RE &  \% DE & DE only & DW    & MBPT \\
		\hline\hline
		3$s$  & 710 - 1120 & 110.6 $\pm$ 5.7 & 108.29 & 60.82 & 60.35 & 43.8  & 56.2  & -0.8  & 2.2   & 2.6 \\
		3$d$  & 830 - 1120 & 47.6 $\pm$ 4.2 & 57.19 & 52.69 & 47.56 & 7.9   & 92.1  & -10.8 & -20.1 & -9.4 \\
		\hline
		3$s$/3$d$ & 830 - 1120 & 1.64 $\pm$ 0.17 & 1.35  & 1.02  & 1.11  & 24.6  & 75.4  & 8.0   & 17.5  & 10.6 \\
		\hline\hline
	\end{tabular}%
	\label{tab:res}%
\end{table*}%
%

%-----------------------------------------------------------------------------
%
\section{Results and discussions}
%
%-----------------------------------------------------------------------------
% 

Overall, we observe an agreement over a wide range of electron energies. However, a few discrepancies are noticeable. The comparison of TSR DR data with the present experiment in Fig.~\ref{fig:3d}a shows minor differences in the baseline, since for clarity the contributions to the total cross section from $n=4-2$ photon emission that are visible in the 2D plot at $\sim$1~keV photon energies (see~Fig.~\ref{fig:spectro}) were left out, while they are present in the TSR data. The TSR data do not include DE and RE processes, thus its signal disappears above the threshold at ~0.8~keV.

Another discrepancy in Fig.~\ref{fig:3s} at electron energies of 0.72~keV could not be attributed to the $3s$ ROI selection or contamination by a DR feature at~0.83~keV photon energy. 
Moreover, Fe~\textsc{xvii} ($2p^5 3s nl$) DR channels at similar electron energy also predict negligible contributions at $\sim$0.7~keV photon energy.  
Calculations assuming $\sim$10\% of Fe~\textsc{xvi}, which is five times the predicted concentration based on comparison of upward and downward scans, still predict a negligible contribution of Fe~\textsc{xvi} at this energy.

We also checked possible contaminations due to oxygen ions producing~O~\textsc{viii}~Ly$\beta$ and~Ly$\gamma$ lines~\citep{bbb2002,glt2011}. 
These contributions can be estimated by checking the presence of $KLL$ DR resonances of~O~\textsc{vii-v} in the electron-energy range of 0.42--0.55~keV, which is free from low-energy contamination from Fe~\textsc{xvii} DR resonances, see Fig.~\ref{fig:spectro}. 
We found out less than 1\% $K\alpha$ X-ray counts after correcting for detector efficiency, which could be due to O~\textsc{viii-iii} $KLL$ DR, compared to the strong Fe $LMM$ features. 
Moreover, simultaneous measurements with vacuum-ultraviolet (VUV) spectrometer in the 40--140 nm wavelength range, which we also used to monitor impurity ions, did not show any fluorescent lines of O~\textsc{iv-v} as well as strong fine-structure $2s-2p$ transitions of O~\textsc{vi}.
This clear absence of any fluorescent X-ray or VUV lines essentially shows that there are no oxygen ions co-trapped with Fe ions in our EBIT, and rules out O~\textsc{viii}~Ly$\beta$ and~Ly$\gamma$ lines as source systematics in our measurements. 
Thus, oxygen impurity lines can be disregarded as a cause of observed discrepancies between the present experiment and theory.

We further investigated the contribution of charge-exchange recombination X-rays in our experiment. 
Charge exchange (CX) can also produce Fe~\textsc{xvi} ions by recombining an electron from residual gases with Fe~\textsc{xvii} ions.
However, X-rays due to CX do not directly affect Fe~\textsc{xvii} $n=3-2$ transitions. It only indirectly affects the emission due to the large fraction of Fe~\textsc{xvi} ions which may produce satellites due to the inner-shell excitation. 
A comparison between $LMM$ DR scans above and below the production threshold of Fe~\textsc{xvii} shows only a negligible contribution of Fe~\textsc{xvi} ions in our trap. 
CX into Fe~\textsc{xviii} producing Fe~\textsc{xvii} can also be neglected since the upper electron beam energy scan limit ($\sim$1.12 keV) is below the production threshold of Fe~\textsc{xviii} ions ($\sim$1.27 keV). 
Moreover, we used a four-stage differential pumping system to inject the iron pentacarbonyl compound. In the present experiment, the injection pressure at the second stage was set too low (~$\lesssim8\times10^{-9}$), which follows two additional stages of cryogenic differential pumping at 45 K and 4 K temperature shield apertures at the Heidelberg FLASH-EBIT~\citep{epp2010}. 
Due to this, at the vacuum level well below~$\lesssim1\times10^{-11}$ mbar in the trap region, CX rates reduce by three orders of magnitudes. 
Our previous CX measurements with highly charged S, Ar, and Fe ions show that additional two-to-three orders of magnitude raise in in the injection pressure is required to observed CX X-ray features~\citep{shah2016}.  
Therefore, CX can be neglected as a cause of these discrepancies.

The $3s$ emission in Fig.~\ref{fig:3s} shows a dominant contribution of RE in the proximity of the excitation threshold (0.73--0.8~keV) of the total photon emission. 
This can be traced back to the $LMP$ ($2p^5 3l 6l'$) and $LNN$ ($2p^5 4l 4l'$) resonances. 
Previously,~\citet{pindzola2006} predicted the $LMP$ RE contribution to the $2p^5 3s$ DE cross section at $\approx$750~eV. 
In addition to this, we observe RE channels up to $LPP$ ($2l 6l'6l''$), which are necessary to fully understand our data up to $E\approx$0.11~keV. 
Furthermore, DE processes populate states having non-dipole (e.~g.~ $2p^5 3p$) decays to the ground state, feeding the $3s$ manifold through radiative cascades. 
The excitation thresholds of such states are apparent in the stepwise shape of the non-RE theoretical prediction in Fig.~\ref{fig:3s}. 
Excellent agreement is observed when all these atomic processes are taken into account and compared with both DW and MBPT theoretical methods. 
Table~\ref{tab:res} lists theoretical and experimental total cross sections, as well as the contribution of RE and DE, showing detailed agreement between both theories and experiment for the $3s$ manifold.

On the other hand, the $3d$ emission shown in Fig.~\ref{fig:3d}b and its total cross section in Table~\ref{tab:res} demonstrate an overestimation of DW theory by 20\%, while MBPT overpredicts by only 9\%. 
Such differences also appear in earlier predictions~\citep{cpr2002,lpb2006,che2007,chen2008,chen2011}. 
For example, previous predictions for the $3d$ cross sections reported by~\citet{lpb2006} and~\citet{cpr2002} differ by 14\% and 17\% from our measurements, respectively.  
The difference between our present MBPT and experiment is likely due to the $3C$ component of the $3d$ manifold (see Fig.~7 of~\citet{gu2009}). 
As discussed by~\citet{che2007, chen2008mnras, gu2009, chen2011, slt2015}, the main difference between theoretical methods evaluating the $3C$ component can be traced back to the completeness of electron correlation, and not to the theoretical scattering implementation, i.~e.,~regardless of using DW, close-coupling, or R-matrix methods. 
The MBPT-corrected cross sections presented here give a better agreement with the measurements because of their more complete treatment of correlation effects in comparison with our DW cross sections. 
We note that the remaining discrepancy on the $3C$ component indicates that higher-order correlation effects beyond second order need to be considered~\citep{safronova2001,slt2015}, as also confirmed in the earlier findings~\citep{sbr2012}. 
Laboratory measurements~\citep{bbc2006} performed at the LLNL-EBIT are within our uncertainties (see~Fig.~\ref{fig:3d}), and share the same conclusion that the actual $3C$ cross sections are lower than theoretical predictions. 
Here, we note that the present measurement resolved RE components for the $3d$ manifold and found their overall contributions to be $\sim$8\%, which is also consistent with values reported by~\citet{bbc2006}.

The ratios of $3s/3d$, which are \textit{independent} of cross-section normalization and are of astrophysical preeminence, plotted in Fig.~\ref{fig:ratio} as a function of electron beam energies, are also in good agreement with the LLNL-EBIT results of~\citet{bbb2002}. 
We found $\sim$18\% discrepancy in $3s/3d$ ratios between our results and DW predictions. Similarly to the previous reasoning, including a more complete treatment of electron correlation for the $3C$ through the MBPT method reduces these discrepancies to $\sim$11\%. 
The Fig.~\ref{fig:ratio} also demonstrate the substantial contribution of RE ($\sim$25\%); thus, it is important to take into account RE in evaluating $3s/3d$ ratios.

%
%
%-----------------------------------------------------------------------------
%
\section{Summary and conclusions}
%
%-----------------------------------------------------------------------------
%
In this work, we reported a first systematic measurement of the soft X-ray emission in~Fe~\textsc{xvii} after electron recombination with six-fold improved electron energy resolution and three-fold improved counting statistics for both the $3s$ and $3d$ line manifolds.
In contrast to previous work where excitation cross sections measured only for a few electron energies, we measured Fe~\textsc{xvii} line emission cross sections for continuous electron beam energies from 0.3 to 1.1 keV. 
We demonstrated that predictions have to include substantially larger sets of configurations and the corresponding contributions of DR and RE to the main DE process, as well as the effect of forbidden transitions driving the line emission through radiative cascades. 
Our DW and MBPT predictions show a good agreement with the measured $3s$ cross sections.
On the other hand, the discrepancy for the $3d$ excitation and for $3s/3d$ ratios found in earlier studies were broadly confirmed in this work. 
The agreement between our measurements and previous laboratory data based on storage ring, ion trap, and tokamak measurements, and the reproducibility of the observed discrepancy in the $3d$ manifold, let us conclude in accord with earlier works that the main disparities between models and astrophysical observations in the $3s/3d$, $3C/3D$, and $3C/(3G+M2)$ are due to an overestimation of the $3C$ component.

Our experimental data and dedicated calculations may contribute to a better understanding of these line ratios, which are used for estimating opacity and turbulence velocities in galaxies~\citep{pzw2012,bbg2004}. 
Moreover, measured DR satellites associated with Fe~\textsc{xvii} can provide an excellent diagnostics of coronal temperatures of stars~\citep{beiersdorfer2018}.   
Our results can be employed to benchmark widely-used spectral codes, such as \textsc{SPEX}~\citep{kaastra1996,gu2019},~\textsc{AtomDB}~\citep{fsb2012}, and~\textsc{CHIANTI}~\citep{Dere_2019}. 
Validation of such codes is, indeed, an urgent task in view of the upcoming launch of the X-ray microcalorimeter-based~\textit{XRISM} satellite~\citep{xrism2018,hitomi2018,labastrowhite2019}. 
The expected scientific harvest of this mission, and the future ones~\textit{Arcus}~\citep{arcus} and~\textit{Athena}~\citep{barret2016} should revolutionize X-ray astrophysics, as the few but nonetheless, excellent measurements of the ill-fated \textit{Hitomi} mission have shown~\citep{hitomi2016,hitomi2017}.

%
%-----------------------------------------------------------------------------
%
\section{Acknowledgements}
%
%-----------------------------------------------------------------------------
We acknowledge Prof. Dr. Stefan Schippers for providing the raw data of dielectronic recombination rates measured at the Test Storage Ring. We also thank Dr. Zolt\'an Harman for valuable discussion on this work. P.~A. acknowledges the support from Funda\c{c}\~{a}o para a Ci\^{e}ncia e a Tecnologia (FCT), Portugal, under Grant No.~UID/FIS/04559/2013(LIBPhys) and under Contract No.~SFRH/BPD/92329/2013. This work was also supported by the Deutsche Forschungsgemeinschaft (DFG) Project No.~266229290.

%
%-----------------------------------------------------------------------------
%
%{References}
%
%-----------------------------------------------------------------------------
%
\bibliographystyle{yahapj}

\end{document}